\documentclass[a4paper]{jpconf}
\usepackage{citesort}
\usepackage{iwsmihead}
\usepackage{amsmath}
\usepackage{graphicx}

\newcommand{\vz} {{\bf z}}

\newcommand{\vmF} {{\bf \mathcal{F}}}
\newcommand{\vmG} {{\bf \mathcal{G}}}
\newcommand{\vx}  {{\bf x}}

\newcommand{\sU} {{\tilde{U}}}
\newcommand{\sr} {{\tilde{r}}}
\newcommand{\iintinf} {{\int^{\infty}_{-\infty}\int^{\infty}_{-\infty}}}

\begin{document}
\title{Tracking Dynamics 
of Two-Dimensional Continuous Attractor Neural Networks}

\author{C. C. Alan Fung$^1$, K. Y. Michael Wong$^1$ and Si Wu$^2$}

\address{
$^{1}$Department of Physics, The Hong Kong University of Science and
Technology, Hong Kong, China\\
$^2$Lab of Neural Information Processing, Institute of Neuroscience,
Shanghai, China }

\ead{alanfung@ust.hk, phkywong@ust.hk, siwu@ion.ac.cn}

\begin{abstract}
We introduce an analytically solvable model of two-dimensional
continuous attractor neural networks (CANNs). The synaptic input and
the neuronal response form Gaussian bumps in the absence of external
stimuli, and enable the network to track external stimuli by its
translational displacement in the two-dimensional space. Basis
functions of the two-dimensional quantum harmonic oscillator in
polar coordinates are introduced to describe the distortion modes of
the Gaussian bump. The perturbative method is applied to analyze its
dynamics. Testing the method by considering the network behavior
when the external stimulus abruptly changes its position, we obtain
results of the reaction time and the amplitudes of various
distortion modes, with excellent agreement with simulation results.
\end{abstract}

\section{Introduction}
Continuous attractor neural networks (CANNs) are very useful models
for describing the encoding of continuous stimuli in neural systems
\cite{Amari77,Ermentrout98,Seung96,Wu08,Zhang96}. The encoded
stimuli can either be some simple features of objects, such as their
orientations \cite{Ben-Yishai95}, moving directions
\cite{Georgopoulos93} and spatial locations \cite{Samsonovich97}, or
some complicated rules that underly the categorization of objects
\cite{Jastorff06}. Compared with other attractor models, CANNs have
the distinctive feature that they hold a family of stationary states
which can be translated into each other without the need to overcome
energy barriers. In the continuum limit, these stationary states
form a continuous manifold in which the system is neutrally stable,
and the network state can translate easily when the external
stimulus changes continuously. Beyond pure memory retrieval, this
large-scale structure of the state space endows the neural system
with a tracking capacity.

To construct a model for CANN, 
the key is that the neuronal interactions should be
properly balanced in excitation and inhibition (e.g., of the
Mexican-hat shape) and be translationally invariant. The former
enables the network to have a local persistent bump solution and the
latter ensures that the network has a continuous family of such
solutions. Although mathematically it is possible to construct a
CANN of dimensionality larger than two, the research on CANNs has so
far been mainly focused on one or two dimensional case. This is
because in the cortex, neurons are essentially distributed in a
two-dimensional sheet. To maintain a CANN of dimensionality larger
than two, it is difficult to wire neurons in a two-dimensional space
without their signals interfering with each other. To encode
continuous features of high dimensionality, the brain may employ
layers of neurons to combine low-dimensional CANNs hierarchically.

The tracking dynamics of a CANN has been theoretically investigated
by several authors in the literature
\cite{Ben-Yishai95,Folias04,Hansel98,Wu05,Zhang96}. These studies
have demonstrated that a CANN has the capacity of tracking a moving
stimulus continuously and that this tracking behavior can describe
many brain functions well. Despite these successes, however,
detailed rigorous analysis of tracking behaviors of a CANN is still
lacking. In a recent work \cite{Fung08,Fung09}, the authors have
developed a perturbative approach to elucidate the tracking
performance of a one-dimensional CANN clearly. Because of the bump
shape of the stationary states of the network, we used the wave
functions of the quantum harmonic oscillator as the basis to decompose
the network dynamics into different motion modes. These modes have
clear physical meanings, corresponding to distortions in the
amplitude, position, width or skewness of the network state. Due to
the neutral stability of network states, the dynamics of a CANN is
typically dominated by a few motion modes, with their contributions
determined by the corresponding eigenvalues. We therefore can
project the network dynamics on these dominating modes and simplify
the network dynamics significantly. In this study, we extend the
perturbative approach to a two-dimensional CANN. The two-dimensional
CANN has much richer dynamics and distortion patterns than the
one-dimensional case \cite{Coombes05,Taylor99,Werner01}. To
elucidate the effect of different distortion patterns on the network
dynamics clearly, we develop the perturbative approach in both
rectangular and polar coordinates. To test our method, we study the
tracking performance of the network when the external stimulus
position experiences an abrupt change. Simulation results confirm
that our method works very well.

\section{The Model}
We consider a two-dimensional neural network coding the stimulus
$\vx = (x_1, x_2)$, with $N$ neurons distributed over this space.
For simplicity, the neurons are assumed to be uniformly distributed
in the space. Considering the common case that the range of possible
values of the stimulus being much larger than the range of neuronal
interactions, we can effectively take $-\infty<x_1, x_2<\infty$. The
dynamics of the synaptic input $U(\vx,t)$ and neuronal response
$r(\vx,t)$ is given by
\begin{eqnarray}
  \tau \frac{\partial U(\vx,t)}{\partial t} & = &
  I^{\rm ext}(\vx) + \rho \iintinf d\vx' J(\vx,\vx')r(\vx')
  - U(\vx, t);
  \label{eq:dUdt2D} \\
  r(\vx, t) & = & \frac{U(\vx,t)^2}{1+k\rho \iintinf d\vx' U(\vx',t)^2},
  \label{eq:r2D}
\end{eqnarray}
where $J(\vx, \vx')$ is the translationally
invariant coupling function defined by
\begin{equation}
  J(\vx, \vx') = \frac{A}{2\pi a^2}
  \exp \left[ -\frac{|\vx-\vx'|^2}{2a^2} \right],
\label{eq:J2D}
\end{equation}
$a$ is the tunning width of the neural network, $k$ is the global
inhibition, and $\rho$ is the density of neurons over the space.
When $I^{\rm ext} =0$ and $0 < k < k_c \equiv A^2\rho/(32\pi a^2 )$,
we have the steady solutions, or {\it stationary states}, given by
(see figure 1)
\begin{eqnarray}
  \sU(\vx|\vz) & = & U_0 \exp \left[ -\frac{|\vx-\vz|^2}{4a^2} \right],
  \label{eq:sU} \\
  \sr(\vx|\vz) & = & r_0 \exp \left[ -\frac{|\vx-\vz|^2}{2a^2} \right],
  \label{eq:sr}
\end{eqnarray}
where
$U_0 = [1 + (1 - k/k_c )^{1/2} ]A/(8\pi a^2 k)$ and
$r_0 = [1 + (1 - k/k_c )^{1/2} ]/(4\pi a^2 k\rho)$.
It is notable that Eqs.~(\ref{eq:sU}) and (\ref{eq:sr})
are valid for any $\vz$.
For simplicity, we consider
$I^{\rm ext} = \alpha U_0 \exp[-|\vx - \vz_0|^2/(4a^2)]$,
where $\alpha$ is the strength of the stimulus.
Thanks to the translational invariance of the coupling function,
the stationary state solution can be peaked at any point in the space.
In this paper, we consider the network response
to a stimulus abruptly changed from $(x_1,x_2)=(0,0)$
to $(z_{01},z_{02})$ at $t=0$.
As shown in the simulation result in figure 2,
the synaptic input can track the change.

\begin{figure}[ht]
\centering
\begin{minipage}{14pc}
\includegraphics[width=14pc]{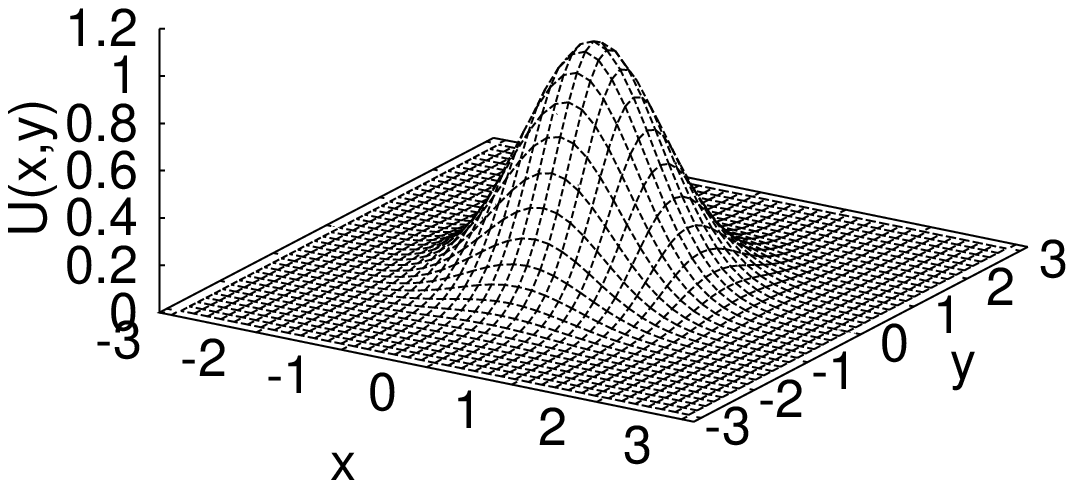}
\caption{\label{fig:sU}A stationary state $\sU(\vx,0)$.
Parameters: $N=40\times 40$, $a=0.5$, $k = 0.5$,
$\tau = 1$ and $\rho = N/(2\pi)^2$.}
\end{minipage}\hspace{2pc}%
\begin{minipage}{14pc}
\includegraphics[width=14pc]{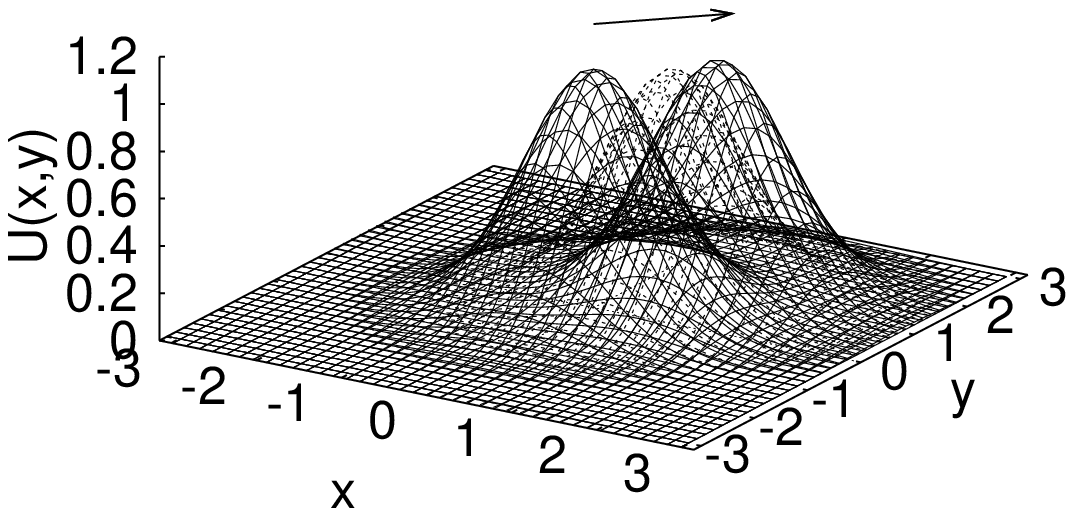}
\caption{\label{fig:track}
The synaptic input in the space due to the abruptly changed stimulus.
Parameters: $\alpha = 0.05$, $\vz_0(t<0) = 0$, $\vz_0(t\ge 0) = (1, 1)$ 
and the rest are the same as figure 1.}
\end{minipage}
\end{figure}

\section{Solution to the Model}
Under the driving of an external stimulus, the network state (i.e,
the bump) moves from its initial position to the target one, with
its shape distorted during the tracking process. Thus, to describe
the tracking performance of a CANN, the key is to know the
distortion patterns and their effects on the network dynamics. We
denote the the state distortion to be $\delta U(\vx,t)\equiv
U(\vx,t)-\tilde U(\vx|\vz)$, whose dynamics is given by 
linearizing Eq.(1) at $\tilde U(\vx|\vz)$ \cite{Fung08,Fung09}, 
\begin{equation}
    \tau\frac{\partial}{\partial t} \delta U(\vx,t)
    =\int^\infty_{-\infty}\int^\infty_{-\infty} d\vx'F(\vx,\vx'|\vz)\delta U(\vx',t)
    -\delta U(\vx,t),
\label{eq:fluc}
\end{equation}
where the interaction kernel $F(\vx,\vx'|\vz)$ is
\begin{equation}
    F(\vx,\vx'|\vz)=\frac{2\rho U(\vx')}{B}
    \left[J(\vx,\vx')-k\rho\int^\infty_{-\infty}\int^\infty_{-\infty} d\vx'' J(\vx,\vx'')r(\vx'')\right],
\label{eq:kernel}
\end{equation}
where $B = 1+k\rho \iintinf \sU(\vx'|\vz)^2 d\vx'$ is a constant.
The network dynamics is determined by the eigenfunctions and
eigenvalues of the kernel $F(\vx,\vx'|\vz)$. To compute them, we
choose the eigenfunctions of the quantum harmonic oscillator as
the basis. While the results are more clearly presented using basis
functions in polar coordinates, the analysis is more conveniently
done in the rectangular coordinates. Hence we will describe the
basis functions in both coordinates.

\subsection{Using Basis Functions in Rectangular Coordinates}

Under the rectangular coordinates, the basis functions are
\begin{equation}
  u_{n,m}(\vx|\vz)=
  \sqrt{\frac{1}{2\pi a^2 m!n!2^{n+m}}}
  H_n\left(\frac{x_1-z_1}{\sqrt{2}a}\right)
  H_m\left(\frac{x_2-z_2}{\sqrt{2}a}\right)
  \exp \left[-\frac{|\vx-\vz|^2}{4a^2}\right],
\label{eq:unm}
\end{equation}
where $H_n(x)$ is the $n^{th}$ order Hermite Polynomial \cite{Griffiths04}.
By the completeness of the basis functions,
we have
\begin{equation}
  U(\vx, t)
  = \sU(\vx|\vz(t)) + \sum_{n,m} a_{n,m}(t) u_{n,m}(\vx|\vz(t)).
  \label{eq:dU}
\end{equation}
Thanks to the orthonormality of the basis functions, we have derived
\cite{Fung08,Fung09}
\begin{eqnarray}
  \tau \frac{da_{n,m}}{dt}
  & = & I_{n,m} + \sum_{n',m'}\vmF_{n,m,n',m'}a_{n',m'}(t) - a_{n,m}
  \nonumber \\
  &~& -\frac{\tau}{2a}\frac{dz_1}{dt}
  \left[\sqrt{2\pi}aU_0\delta_{n1,m0} + \sqrt{n} a_{n-1,m}(t)
  - \sqrt{n+1} a_{n+1,m}(t)\right]
  \nonumber \\
  &~& -\frac{\tau}{2a}\frac{dz_2}{dt}
  \left[\sqrt{2\pi}aU_0\delta_{n0,m1} + \sqrt{m} a_{n,m-1}(t)
  - \sqrt{m+1} a_{n,m+1}(t)\right],
\label{eq:danm}
\end{eqnarray}
where $I_{n,m}$ is the projection of $I^{\rm ext}$ onto
$u_{n,m}$, given by
\begin{equation}
  I_{n,m} (\vx|\vz_0) = \alpha U_0 \sqrt{\frac{2\pi a^2}{n!m!}}
  \left(\frac{z_{01} - z_1}{2a}\right)^n
  \left(\frac{z_{02} - z_2}{2a}\right)^m
  \exp \left[-\frac{|\vz-\vz_0|^2}{8a^2}\right],
\label{eq:Inm}
\end{equation}
and $\vmF_{n,m,n',m'}$, the interaction matrix, is given by
\begin{eqnarray}
  \vmF_{0,0,0,0} & = & 1-\left(1-\frac{k}{k_c}\right)^{\frac{1}{2}}, 
\label{eq:F1}\\
  \vmF_{n,m,n'm'} & = &\sqrt{\frac{m'!n'!}{m!n!}}\frac{2}{2^{m'+n'}}
  \frac{(-)^{\frac{m'+n'-m-n}{2}}}{2^{\frac{m'+n'-m-n}{2}} (\frac{m'-m}{2})!
  (\frac{n'-n}{2})!}, \nonumber \\
  &~&~~~~~~~~~~~~~~~~~~~~~~{\rm~~if~} \tfrac{n'-n}{2}{\rm~and~}\tfrac{m'-m}{2}
       {\rm~are~positive~integers}\label{eq:F2}\\
  \vmF_{n,m,n'm'} & = & 0,
  ~~~~~~~~~~~~~~~~~~~~ {\rm~~otherwise}.\label{eq:F3}
\end{eqnarray}
The center of mass is given by the self-consistent condition
$\vz(t) = \iint d\vx U(\vx,t)\vx / \iint d\vx U(\vx,t)$. 
If the external stimulus is symmetric with respect to the $x$-axis, 
then $dz_2/dt=0$ and 
\begin{equation}
  \frac{dz_1}{dt}
  = \frac{2a}{\tau}
  \frac{\sum_{{\rm odd}~n,~{\rm even~} m}
      \sqrt{\frac{(m-1)!!}{m!!}\frac{n!!}{(n-1)!!}}
      \left(I_{n,m} + \sum_{n',m'}\vmF_{n,m,n',m'}a_{n',m'} \right)
  }{\sqrt{2\pi}aU_0 +
    \sum_{{\rm even}~n,~m}\sqrt{\frac{(m-1)!!}{m!!}\frac{(n-1)!!}{n!!}} 
  a_{n,m}}.
\label{eq:dzrec}
\end{equation}
The eigenvalues of $\vmF$ are
$\lambda_{0,0} = 1- (1-k/k_c)^{1/2}$,
$\lambda_{n,0}=\lambda_n$ for $n\neq 0$,
$\lambda_{0,m}=\lambda_m$ for $m\neq 0$,
$\lambda_{n,m} = \lambda_n\lambda_m$ for $n\neq0$ and $m\neq0$,
where $\lambda_n=2/2^n$.
From this result, one can conclude that, if the stimulus is absent,
all modes of distortion will decay exponentially in time,
except for the eigenfunctions $u_{1,0}$ and $u_{0,1}$,
whose eigenvalues are 1.  To prove this, we define $v^R_{n,m}(\vx|\vz)$ 
to be the right eigenfunctions of $\vmF$. 
Then we may express $\delta U(\vx,t)$ in
Eq.~(\ref{eq:fluc}) as $\delta U(\vx,t) = \sum_{n,m} \delta U_{n,m}(\vz,t)
v^R_{n,m}(\vx|\vz)$.  Using the orthonormality of the left and 
right eigenfunctions of $F(\vx,\vx'|\vz)$, we have
\begin{equation}
    \delta U_{n,m}(z,t)=\iintinf d\vx\delta U(\vx,t)v^L_{n,m}(\vx|\vz).
\end{equation}
Assume that the motion of the bump is slow,
so that $d\vz/dt$ becomes negligible in Eq.~(\ref{eq:fluc});
as we shall see,
this assumption is valid as long as the external stimulus is sufficiently weak.
Then, the projection of Eq.~(\ref{eq:fluc}) on the eigenfunctions become
\begin{equation}
    \tau\frac{d}{dt}\delta U_{n,m}(\vz,t)=(\lambda_{n,m}-1)\delta U_{n,m}(z,t).
\label{eq:dUnm}
\end{equation}
Hence,
\begin{equation}
    \delta U_{n,m}(\vz,t)=\delta U_{n,m}(\vz,0)
    \exp\left[-{\frac {(1-\lambda_{n,m})t} {\tau}}\right],
\label{eq:Unm}
\end{equation}
where $\delta U_{n,m}(\vz, 0)$ is the initial value of the projection.  
$u_{1,0}$ and $u_{0,1}$ correspond 
to the trackability of the synaptic input as well as the
neuronal response, because $u_{1,0}\sim \partial \sU /\partial x_1$ and
$u_{0,1}\sim \partial \sU /\partial x_2$. $u_{1,0}$ and $u_{0,1}$
are the modes of the position shift of the synaptic input.
Thus, it guarantees the stability of the stationary solution and
trackability of the synaptic input.

We are now ready to find the tracking solution
to the stimulus abruptly changed from $(0,0)$ to $(z_{01},0)$ at $t=0$.
Neglecting the depdendence on all $a_{n,m}$ terms in Eq.~(\ref{eq:dzrec}), 
we have
\begin{equation}
  \frac{dz_1}{dt}
  = \frac{\alpha}{\tau} (z_{01} - z_1)
  \exp\left[-\frac{(z_{01}-z_1)^2}{8a^2}\right],
\label{eq:rec:l1}
\end{equation}
which is consistent with the result obtained
from the one-dimensional case \cite{Fung08,Fung09}.
This approximation is useful when $|\vz-\vz_0|$ is small
and the stimulus is weak.
It also shows that the tracking behavior is similar
to the one-dimensional case.

\subsection{Using Basis Functions in Polar Coordinates}

Since the system is also rotationally invariant,
the analysis can proceed
by using the eigenfunctions with polar coordinates.
The eigenfunctions of quantum harmonic oscillators are
\begin{equation}
  \psi_{l,j}(r,\theta) =
  \sqrt{\frac{(\frac{l-j}{2})!(\frac{l+j}{2})!}{2\pi a}}
  \left[ \sum^{\frac{l-|j|}{2}}_{t=0}
    \frac{(-1)^t(\frac{r}{\sqrt{2}a})}{(\frac{l-j}{2})!(\frac{l+j}{2})!t!}
     \right] e^{-\frac{r^2}{4a^2} + ij\theta},
\label{eq:psilj}
\end{equation}
where $i = \sqrt{-1}$,
and $l$, $j$ are the radial and angular quantum numbers respectively.
Decomposing the distortion term, we have
$\delta U = \sum_{l,j}b_{l,j}(t)\psi_{l,j}$.
The matrix elements of the transformation matrix are
\begin{equation}
  [T]_{n,m,l,j} \equiv \langle u_{n,m} | \psi_{l,j} \rangle
  = i^m \sqrt{
    \frac{\left(\frac{l+j}{2}\right)!
          \left(\frac{l-j}{2}\right)!}{n!m!2^l}}
    \left[ \sum_{t}(-1)^t\binom{m}{t}\binom{n}{\tfrac{l-j}{2}-t} \right],
\label{eq:trans}
\end{equation}
Similar to Eq.~(\ref{eq:danm}),
there is an interaction kernel $\vmG$
that represents the interaction between different $\psi_{l,j}$,
which can be obtained by $\vmG = T^{-1}\vmF T$.
The first few eigenfunctions of $\vmG$ are
\begin{eqnarray}
  \Psi_{00} & = & \psi_{00} ,\\
  \Psi_{1\pm1} & = & \psi_{1\pm1},\\
  \Psi_{20} & = & \frac{1}{\sqrt{1+(\lambda_{00}-1/2)^2}}
  \left[\psi_{00} + (\lambda_{00}-1/2)\psi_{20} \right] ,\\
  \Psi_{2\pm2} & = & \psi_{2\pm2},\\
  \Psi_{3\pm1} & = & \frac{1}{\sqrt{3^2 + 2}}\left[\sqrt{2}\psi_{1\pm1}
  + 3\psi_{3\pm1} \right],{\rm~and}\\
  \Psi_{3\pm3} & = & \psi_{3\pm3},
\end{eqnarray}
where the indices $l$ and $j$ of $\Psi_{l,j}$ represent the highest
basis function it contains. Their eigenvalues are $\lambda_{00}$, 1,
1/2, 1/2, 1/4, and 1/4 respectively.

As shown in Figs.~3 to 8, 
the eigenfunctions are symmetric with respect to the origin.
The eigenfunctions correspond to different modes of the distortion of the
synaptic input during the motion.  $\Psi_{00}$ corresponds to the
change in height.
It can describe, say, the reduction of the bump height
during the process to catch up with
the new position of the stimulus, as shown in figure 2.
$\Psi_{1,\pm 1}$ can describe the movement of the bump
towards the preferred positions.
$\Psi_{2,0}$ describes not only changes in the height of the bump,
but also changes in the width of the bump during the motion.
$\Psi_{3,\pm 1}$ describes the skewing
of the bump due to the stimulus and other modes.
While the above distortion modes are apparently
extensions of those in the one-dimension case,
the modes $\Psi_{2,\pm 2}$ and $\Psi_{3,\pm 3}$
are unique to the two-dimensional case.
The former corresponds to an elliptical distortion of the bump shape,
and the latter to a three-fold distortion.

\begin{figure}[ht]
\centering
\begin{minipage}{16pc}
\includegraphics[width=16pc]{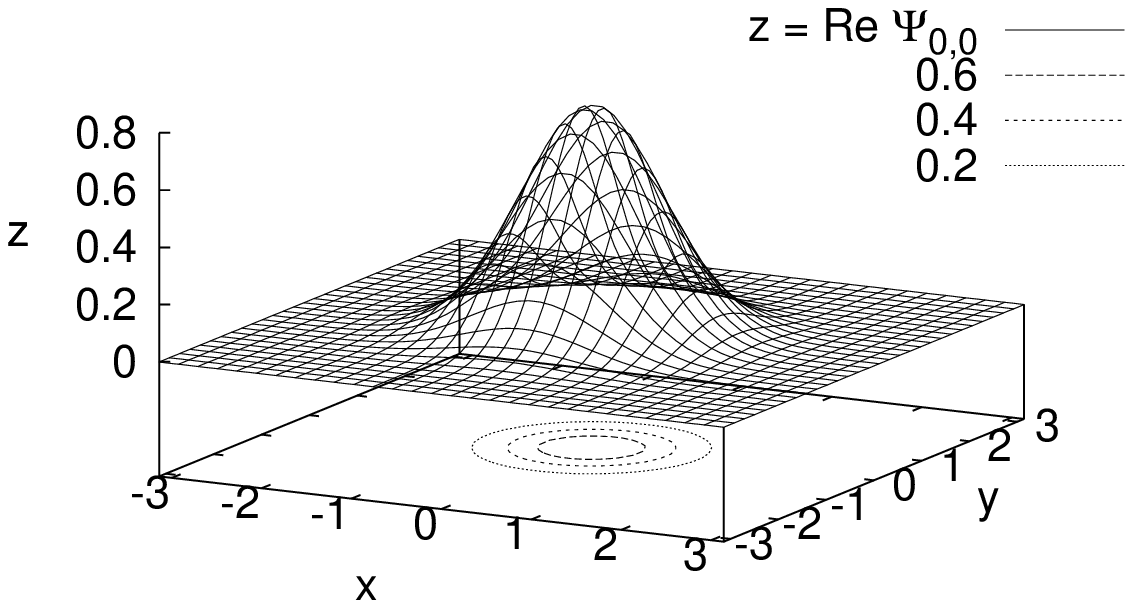}
\caption{\label{fig:Psi00}Real part of the eigenfunction $\Psi_{0,0}$.}
\end{minipage}\hspace{2pc}%
\begin{minipage}{16pc}
\includegraphics[width=16pc]{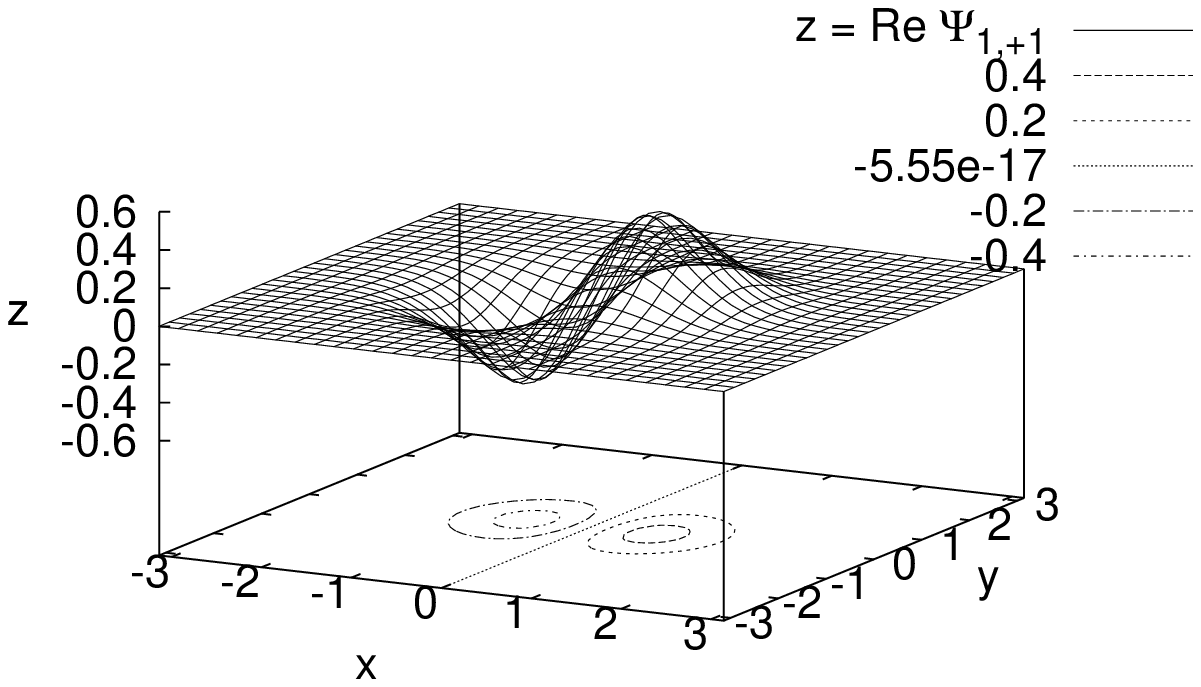}
\caption{\label{fig:Psi11}Real part of the eigenfunction $\Psi_{1,+ 1}$.}
\end{minipage}
\end{figure}

\begin{figure}[ht]
\centering
\begin{minipage}{16pc}
\includegraphics[width=16pc]{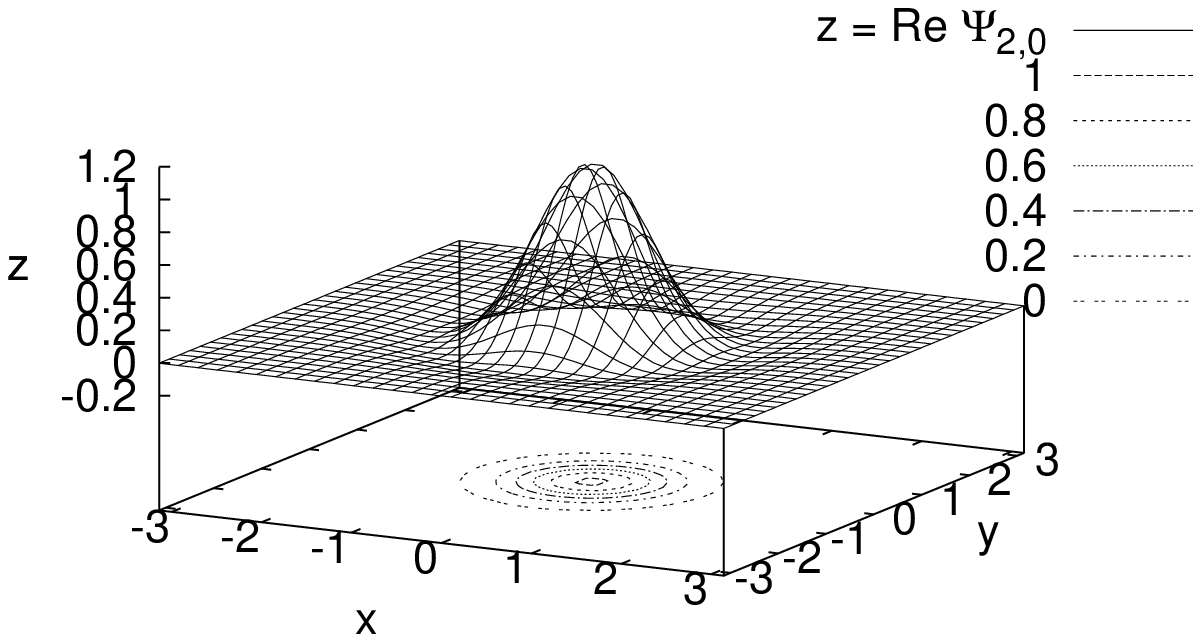}
\caption{\label{fig:Psi20}Real part of the eigenfunction $\Psi_{2,0}$.}
\end{minipage}\hspace{2pc}%
\begin{minipage}{16pc}
\includegraphics[width=16pc]{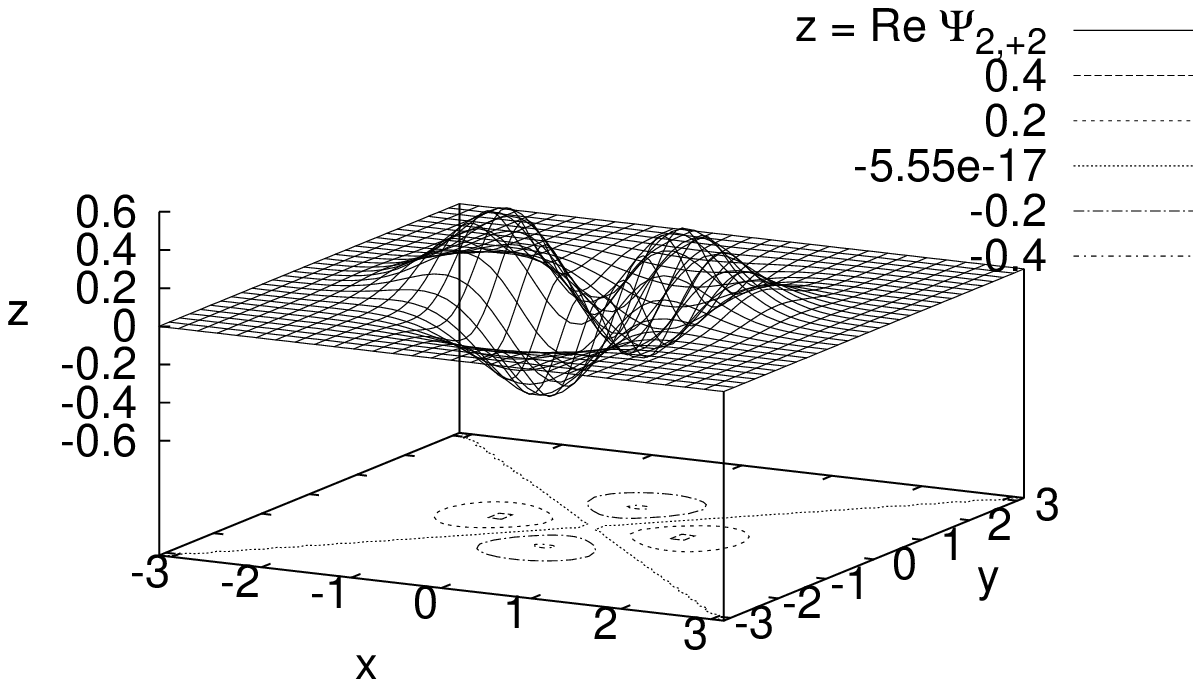}
\caption{\label{fig:Psi22}Real part of the eigenfunction $\Psi_{3,+1}$.}
\end{minipage}
\end{figure}

\begin{figure}[ht]
\centering
\begin{minipage}{16pc}
\includegraphics[width=16pc]{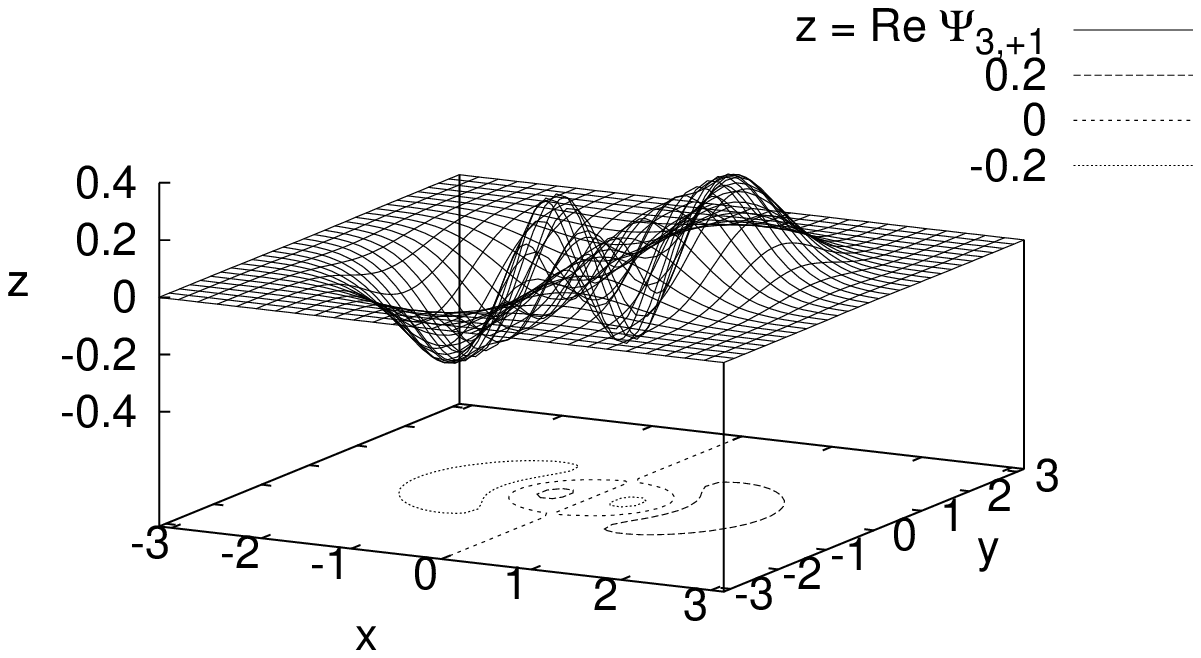}
\caption{\label{fig:Psi31}Real part of the eigenfunction $\Psi_{2,+2}$.}
\end{minipage} \hspace{2pc}%
\begin{minipage}{16pc}
\includegraphics[width=16pc]{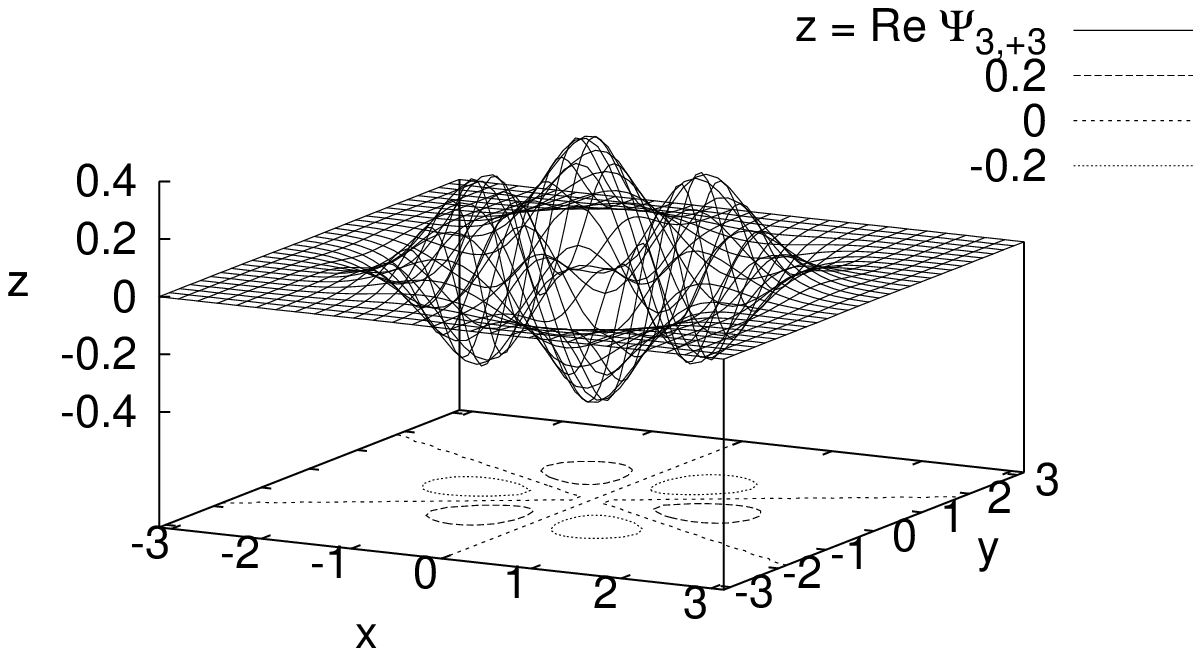}
\caption{\label{fig:Psi33}Real part of the eigenfunction $\Psi_{3,+3}$.}
\end{minipage}
\end{figure}

By using the transformation matrix in Eq.~(\ref{eq:trans}), 
Eq.~(\ref{eq:danm}) can be transformed from rectangular to polar coordinates
up to arbitrary order.
For the perturbation up to $l=2$, 
and for external stimuli symmetric with respect to the $x$-axis, we have

\begin{equation}
  \left(\tau \frac{d}{dt} + 1-\lambda_{00}\right) b_{00}
  = I_{00} - \tau \frac{dR}{dt}
  \left[-\frac{1}{\sqrt{8}a}(b_{1-1} + b_{1+1})\right] - b_{20},
\label{eq:pol:00}
\end{equation}
\begin{equation}
  \tau \frac{d}{dt} b_{1\pm1} =
  I_{1\pm1} - \tau\frac{dR}{dt}\left(
  \frac{\sqrt{2\pi} aU_0 + b_{00}}{2\sqrt{2}a}
  - \frac{1}{2\sqrt{2}a}b_{20} - \frac{1}{2a}b_{2\pm2}\right),
\label{eq:pol:l1}
\end{equation}
\begin{equation}
  \left(\tau \frac{d}{dt} + \frac{1}{2} \right) b_{20}
  = I_{20} - \tau\frac{dR}{dt}\left(-\frac{1}{2\sqrt{2}a}
  (b_{1-1} + b_{1+1})\right),
\label{eq:pol:20}
\end{equation}
\begin{equation}
  \left(\tau \frac{d}{dt} + \frac{1}{2} \right) b_{2\pm2}
  = I_{2\pm2} - \tau\frac{dR}{dt}\frac{1}{2a}b_{1\pm1},
\label{eq:pol:22}
\end{equation}
\begin{equation}
  \tau \frac{dR}{dt} = 2\sqrt{2}a
  \frac{I_{11}}
  {\sqrt{2\pi}aU_0 + b_{00} -\sqrt{2}b_{20} - 2\sqrt{2}b_{22}},
\label{eq:pol:R}
\end{equation}
where $R$ is the radial distance from the origin.
Note that $I_{l,j}=I_{l,-j}$ and $b_{l,j}=b_{l,-j}$
due to the symmetry when the stimulus lies on the $x$-axis.

\section{Simulation Experiments}

In the simulation experiments, the number of neurons is $N = N_x \times N_y$, 
and the range of $(x_1,x_2)$ is $-\pi \leq x_1, x_2 < \pi$.
The boundary condition is periodic.

\subsection{Reaction Time to an Abrupt Change of the Stimulus}

In this experiment, 
the stimulus is centered at $(0,0)$
until the synaptic input $U(\vx,t)$ becomes steady.
At $t=0$, 
the stimulus abruptly changes from $(0,0)$ to $\vz_0 \equiv (z_{01},0)$,
and we observe the dependence of the reaction time on the distance $z_{01}$.
Then, the bump will track the stimulus, as shown in figure 2.
The reaction time is defined by the time needed
to have $|\vz(t) - \vz_0| < \Theta$, where $\Theta$ is the threshold.
This threshold is necessary in this experiment, 
because the motion of $\vz(t)$
will become very slow when it approaches the stimulus,
as implied by Eq.~(\ref{eq:rec:l1}).
Also, the assumption is reasonable because in real biological systems,
we do not need to have $\vz(t) = \vz_0(t)$ to make decisions.

\begin{figure}[ht]
\centering
\vspace{1pc}
\includegraphics[width=16pc]{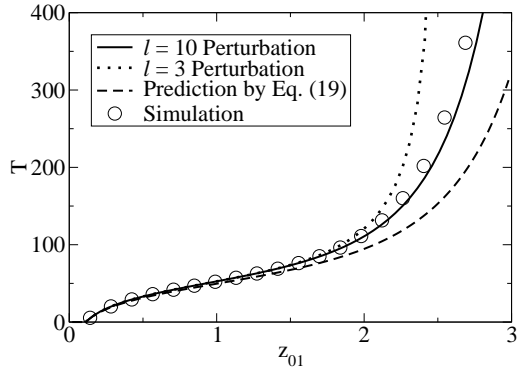}\hspace{2pc}%
\begin{minipage}[b]{14pc}\caption{\label{fig:rtime}
The reaction time $T$ for the
syanptic input to catch up
the stimulus position change from 0 to $z_{01}$.
Parameters: $N= N_x \times N_y = 40\times40$,
$\Theta = \pi\sqrt{2/N}$, $\alpha = 0.05$
and the rest are the same as figure 1.}
\end{minipage}
\end{figure}

As shown in figure 9,
the prediction given by Eq.~(\ref{eq:rec:l1}) works well
only when the change in the position of the stimulus is small,
while the $l=3$ perturbation works well up to $|\vz_0| = 2$.
The prediction of the $l = 10$ perturbation is the best among the three.
From this result, one can state that,
when the position change of the stimulus is small,
only $\psi_{l,j}$ with small $l$ will be activated.
However, when the change in stimulus position is larger,
higher order distortions are activated.
This is reasonable, because, for smaller $l$, the distortions
are concentrated around $\vz(t)$,
but if the stimulus is far away from $\vz(t)$,
the tail part of the bump will be distorted first,
leading to higher order distortions.

\subsection{Amplitudes of the Basis Function Distortion Modes}

From Eqs.~(\ref{eq:danm}) and (\ref{eq:trans}),
$b_{l,j}$ can be predicted
by the projections of $\delta U$ onto the basis functions $\psi_{l,j}$.
The experimental settings are the same as above,
but $\vz_0$ was fixed to be $\vz_0 = (2,0)$ in polar coordinates.

As shown in figure 10, the predicted $b_{l,j}$'s
agree with the simulation results well.
It confirms that the perturbative method can also predict
the motion of the synaptic input in detail.
$b_{0,0}$ indicates that the height drops from its initial value
after the stimulus is shifted.
The $\psi_{0,0}$ component of the distortion is reduced by the inhibition.
It approaches 0 roughly,
as if there were no external stimulus,
and there is even a slight overshoot.
Afterwards, the distortion relaxes smoothly to the equilibrium value
when it approaches the shifted position of the stimulus.
Similarly, the amplitude of the $\psi_{1,\pm 1}$ components
falls abruptly to a negative value initially.
This is due to the tail of the bump
being pulled by the newly positioned stimulus,
causing the peak to lag behind the center of mass.
Afterwards, it relaxes smoothly to 0.
The initial change in $b_{2,0}$
indicates an increase in width along the direction of the stimulus,
and that of $b_{2,\pm 2}$ signals a cigar-shaped distortion
when the bump is being pulled by the stimulus.
The amplitude $b_{3,\pm 1}$ describes the skewness of the bump, 
and $b_{3,\pm 3}$ describes the bump being distorted
when its tip is pulled by the stimulus,
with the posterior part lagging in motion,
causing a triangular-shaped distortion.

\begin{figure}[h]
\centering
\vspace{1.7pc}
\includegraphics[width=28pc]{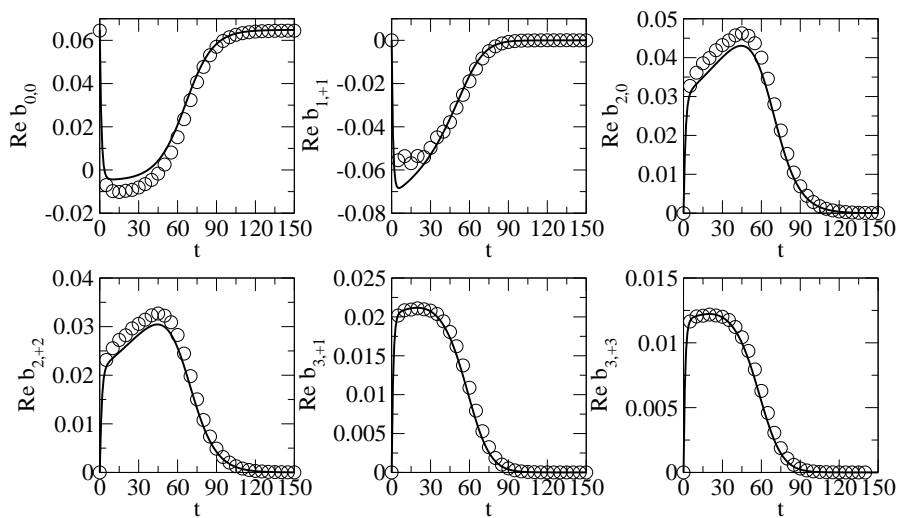}
\caption{\label{fig:bnm}
The experimentally projected $b_{l,j}$ 
and the predicted $b_{l,j}$ using $l=10$
perturbation.  Solid lines: corresponding predictions.
Circles: simulation results.
Parameters: same as figure 9.}
\end{figure}

\section{Conclusion and Discussion}

In this paper, a perturbative method to deal with
continuous attractor neural networks is presented in two-dimensional space.
We have introduced a simple solvable model to demonstrate how to use
perturbative method to analyze the dynamics of distortions and synaptic input.
Since the coupling factors are defined to be translational invariant,
a family of stationary states can be sustained
anywhere in the preferred stimulus space.
Furthermore, the synaptic input is able to
track the stimulus in the space.
By studying the dynamics,
one can deduce the tracking time and the distortions of the synaptic input.

For one-dimensional continuous attractor neural networks,
the tracking speed can be roughly
approximated by
\begin{equation}
    \frac{dz}{dt} =
        \frac{\alpha}{\tau}(z_0-z)\exp\left[-\frac{|z_0-z|^2}{8a^2}\right],
\label{eq:1D:dz}
\end{equation}
which is similar to Eq.~(\ref{eq:rec:l1}). Although they are rough
approximations, the basic properties of tracking in one and two
dimensions are similar. The key predictions in the one-dimensional
case, the maximum trackable speed \cite{Fung08,Hansel98} and the
lag behind a continuously moving stimulus, are also applicable
in two dimensions.  These similarities arise from the rotational 
invariance of the interaction kernel and the bump in the limit
of weak stimulus, since the description of the dynamics by the 
translational mode in sufficient.  Note, However, for the stronger
stimuli, the dynamics is richer in the two-dimensional case, since
distortion modes unique to the two-dimensional case need to be
considered; examples of cigar-shaped and triangular-shaped 
distortions are shown in figure 10.

For this particular model,
the eigenfunctions of the interaction matrix in polar
coordinates are also studied.
By Eq.~(\ref{eq:trans}), the interaction matrix $\vmG$ can be obtained.
However, due to complications in the calculation,
we can only calculate it term by term.
Since the matrix is upper triangular,
the eigenvalues are the diagonal entries, $\lambda_{n,m}$.
As the eigenvalues are at most 1,
one can show that the synaptic input has a stable Gaussian form.
The eigenfunctions corresponding to eigenvalue 1
represent the positional shift.
As studied above, different modes of distortion correspond
to different kinds of distortion.
For example, the component of $\psi_{0,0}$
corresponds to the change in height,
while $\psi_{2,0}$ represents the change in width.

Concerning the robustness of the method, we remark that it can be
applied to other types of networks with tracking behavior. A common
example is the continuous attractor neural network with the Mexican
hat interaction. Using the basis functions of the two-dimensional
quantum harmonic oscillator, we can obtain the matrix elements of
the interaction kernel numerically, although elegant expressions
such as those obtained here may not be available. Perturbation
dynamics can then be worked out analogously. This proposed extension
can be able to address a recent issue of interest, namely, the
instability of bumps and rings in a two-dimensional neural field of
Amari type \cite{Owen07}. Meanwhile, we note in passing that the
present model with a quadratic response and a global inhibition does
not suffer from the stability problem.

\section*{Acknowledgement}
This work is partially supported by 
the Research Grants Council of Hong Kong 
(Grant Nos. HKUST 603607 and HKUST 604008).

\section*{References}

\bibliography{iopart-num}

\providecommand{\newblock}{}
\begin{thebibliography}{10}
\expandafter\ifx\csname url\endcsname\relax
  \def\url#1{{\tt #1}}\fi
\expandafter\ifx\csname urlprefix\endcsname\relax\def\urlprefix{URL }\fi
\providecommand{\eprint}[2][]{\url{#2}}

\bibitem{Amari77}
Amari S 1977 {\em Biol. Cybern.\/} {\bf 27} 77--87

\bibitem{Ermentrout98}
Ermentrout B 1998 {\em Rep. Prog. Phys.\/} {\bf 61} 353--430

\bibitem{Seung96}
Seung H~S 1996 {\em Proc. Acad. Sci. USA\/} {\bf 93} 13339--44

\bibitem{Wu08}
Wu S, Hamaguchi K and Amari S 2008 {\em Neural Comput.\/} {\bf 20} 994--1025

\bibitem{Zhang96}
Zhang K~C 1996 {\em J. Neurosci.\/} {\bf 16} 2112--26

\bibitem{Ben-Yishai95}
Ben-Yishai R, Lev Bar-Or R and Sompolinsky H 1995 {\em Proc. Natl. Acad. Sci.
  USA\/} {\bf 92} 3844--8

\bibitem{Georgopoulos93}
Georgopoulos A~P, Taira M and Lukashin A 1993 {\em Science\/} {\bf 260} 47--52

\bibitem{Samsonovich97}
Samsonovich A and McNaughton B~L 1997 {\em J. Neurosci.\/} {\bf 7} 5900--20

\bibitem{Jastorff06}
Jastorff J, Kourtzi Z and Giese M 2006 {\em J. Vision\/} {\bf 6} 791

\bibitem{Folias04}
Folias E and Bressloff P 2004 {\em SIAM J. Appl. Dyn. Syst.\/} {\bf 3} 378--407

\bibitem{Hansel98}
Hansel D and Sompolinsky H 1998 {\em Methods in Neuronal Modeling: From Ions to
  Networks\/} ed Koch C and Segev I (MIT Press, Cambridge)

\bibitem{Wu05}
Wu S and Amari S 2005 {\em Neural Comput.\/} {\bf 17} 2215--39

\bibitem{Fung08}
Fung C~C~A, Wong K~Y~M and Wu S 2008 {\em Europhys. Lett.\/} {\bf 84} 18002

\bibitem{Fung09}
Fung C~C~A, Wong K~Y~M and Wu S {\em Neural Comput. (in press)\/}

\bibitem{Coombes05}
Coombes S and Owen M~R 2005 {\em Phys. Rev. Lett.\/} {\bf 94} 148102

\bibitem{Taylor99}
Taylor J~G 1999 {\em Biol. Cybern.\/} {\bf 80} 393--409

\bibitem{Werner01}
Werner H and Richter T 2001 {\em Biol. Cybern.\/} {\bf 85} 211--217

\bibitem{Griffiths04}
Griffiths D 2004 {\em Introduction to Quantum Mechanics\/} (Prentice Hall)

\bibitem{Owen07}
Owen M~R, Liang C~R and Coombes S 2007 {\em New J. Phys.\/} {\bf 9} 378

\end{thebibliography}
\end{document}